\documentclass{iau}
\newcommand{\f}{RRAT~J1819--1458}
\usepackage{graphicx}

\def\aj{\ref@jnl{AJ}}                   
             
\def\apj{{ApJ}}                 
\def\apjl{{ApJ}}

\def\mnras{{MNRAS}}

\def\nat{{Nature}}

\title[The extended X--ray emission around RRAT~J1819--1458]{The extended X--ray emission around RRAT~J1819--1458}%% full title %%

\author[A. Camero--Arranz  et al.]  %% short author list %%
{A. Camero--Arranz$^{1}$, N. Rea$^{1}$, M.A. McLaughlin$^{2}$,  N. Bucciantini$^{3,4}$ P. Slane$^{5}$,  
B. Gaensler$^{6}$, D. Torres$^{1,7}$, L. Stella$^{8}$, E. de O\~na$^{1}$, G. Israel$^{8}$, F. Camilo$^{9,10}$ \and  
A. Possenti$^{11}$}

\affiliation{$^{1}$Institut de Ci\`{e}ncies de l'Espai, (IEEC-CSIC), Campus UAB, Fac. de Ci\`{e}ncies, Torre C5, parell, 2a planta, 08193 Barcelona, Spain; email: {\tt camero@ice.cat} \\[\affilskip]
$^{2}$Department of Physics, West Virginia University, Morgantown, WV 26501, USA\\[\affilskip]
$^{3}$INAF - Osservatorio Astrofisico di Arcetri, L.go E.~Fermi 5, 50125, Firenze, Italy\\ [\affilskip]
$^{4}$INFN - Sezione di Firenze, Via G.~Sansone 1, 50019 Sesto Fiorentino, Firenze, Italy\\[\affilskip]
$^{5}$Harvard-Smithsonian Center for Astrophysics, 60 Garden St. Cambridge, MA 02138, USA\\[\affilskip]
$^{6}$The University of Sudney, Room 216, 44 Rosehill Street, Redfern, NSW 2016, Australia\\[\affilskip]
$^{7}$Instituci\'{o} Catalana de Recerca i Estudis Avancats (ICREA)\\[\affilskip]
$^{8}$INAF - Osservatorio Astronomico di Roma, Via Frascati 33, Roma, Italy\\[\affilskip]
$^{9}$Columbia Astrophysics Lab, Columbia University, New York, NY 10027, USA\\[\affilskip]
$^{10}$Arecibo Observatory, HC3 Box 53995, Arecibo, PR 00612, USA\\[\affilskip]
$^{11}$INAF-Osservatorio Astronomico di Cagliari, 09012 Capoterra, Italy
}

\pubyear{2012}
\volume{291}  %% insert here IAU Symposium No.
\jname{\mbox{Neutron Stars and Pulsars: Challenges and Opportunities after 80 years}}
\editors{J. van Leeuwen, ed.}

\begin{document}

\maketitle

%% -- Abstract ----------------------------------
\begin{abstract}
 We present new imaging and spectral analysis of the recently
 discovered extended X--ray emission around the high-magnetic-field
 rotating radio transient RRAT~J1819--1458.  We used two {\it Chandra}
 observations, taken on 2008 May 31 and 2011 May 28. The diffuse
 X--ray emission was detected with a significance  of $\sim$19$\sigma$
 in  the image obtained by combining the two observations. Long-term
 spectral  variability has not been observed.  Possible scenarios for the origin of this diffuse X--ray emission,
 further detailed in  Camero--Arranz et al.\ (2012), 
 are here discussed.
 
\end{abstract}

% add below any authors, subjects and objects for indexing 
%   add more lines if necessary
%   but leave all lines commented out
%\index[author]{LastName1, Initials|textbf}
%\index[author]{LastName2, Initials|textbf}
%\index[subject]{Keyword1}
%\index[subject]{Keyword2}
%\index[object]{Object1}
%\index[object]{Object2}

\firstsection % if your document starts with a section,
\section{Introduction}

Rotating Radio Transients (RRATs) are radio pulsars that were
discovered through their sporadic radio bursts (\cite[McLaughlin et
  al.\ 2006]{mcLaughlin06}). At a radio frequency of 1.4 GHz, radio
bursts are observed from \f ~roughly every $\sim$3 minutes with the
Parkes telescope. The spin period of RRAT~J1819--1458  is 4.3\,s, with
a characteristic age of 117\,kyr at at a 3.6\,kpc distance and a
dipolar magnetic field of  B$\sim$5$\times$10$^{13}$\,G. The spin-down
energy loss rate measured for this source is
\.{E}$_{rot}\sim$3$\times$10$^{32}$ erg s$^{-1}$, being the only
source of this type also detected in  X--rays (\cite[Reynolds et
  al.\ 2006, McLaughlin et al.\ 2007, Rea et al.\ 2008, Kaplan et al. 2009]{reynolds06,mcLaughlin07,rea_cLaughlin08,kaplan09}).  In this work, we present the study of the extended X--ray emission discovered by Rea et al. (2009), resulting from the reduction and combined analysis of two {\it Chandra} observations for RRAT~J1819--1458, performed on 2008 and 2011 May 28.

\section{Observations and data reduction}

The \textit{Chandra} X--ray Observatory observed RRAT~J1819--1458
with the Advanced CCD Imaging Spectrometer (ACIS) instrument on 2008
May 31 (ObsID 7645) for 30\,ks  and again in 2011 May 28 (ObsID 12670)
for 90\,ks,  both in {\tt VERY FAINT} (VF) timed exposure imaging mode. For both observations, we used a 1/8 subarray, which provides a time resolution of 0.4 s.  Standard processing of the data using CIAO software (ver.\,4.4) has been performed.
%, resulting in a final exposure time of 27.88~ks for the first observation and 80.40~ks for the second one.

%%%%%%%%%%%%%%%%%%%%%%%    Image             %%%%%%%%%%%%%%%%%%%%%%%%%%%%%%%%%%

\begin{figure}[t]
\begin{center}
\includegraphics[width=0.95\textwidth]{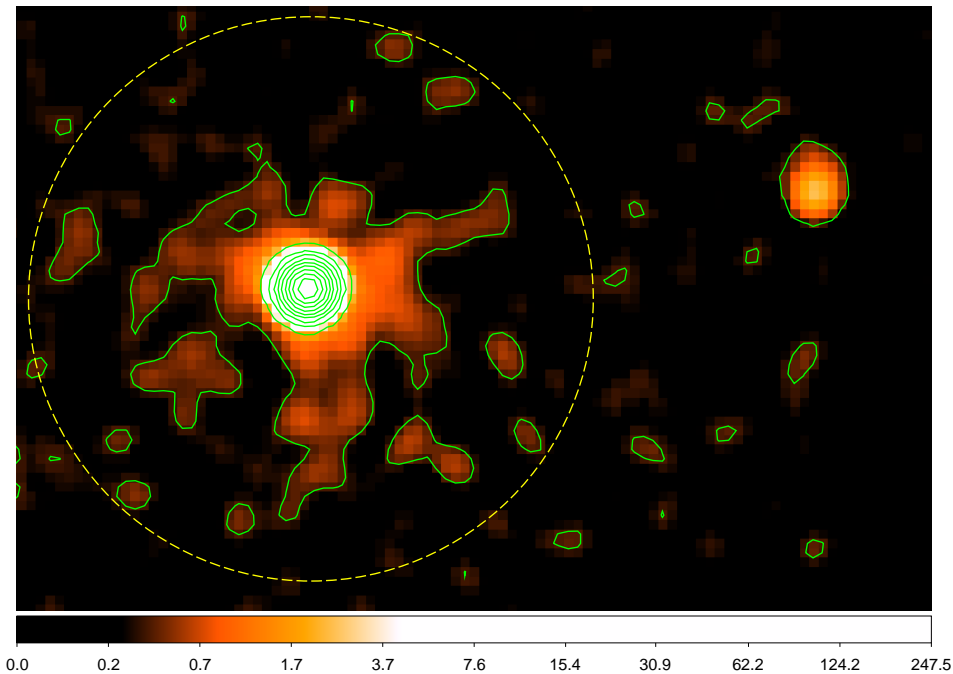} 
\caption{Combined 0.3--10\,keV log image  of RRAT~J1819--1458, with a circular region of 15${"}$ overplotted and the 3$\sigma$ contours of the extended emission (Camero--Arranz et al. 2012). One ACIS-S pixel is  0$"$.492. 
}\label{ima}
\end{center}
\end{figure}
%%%%%%%%%%%%%%%

\section{Analysis and  results}
\subsection{Imaging}

To study the extended X--ray emission found by \cite[Rea et al. (2009)]{rea09} in more detail,  we proceeded with the extraction of a combined image in the 0.3--10 keV energy range, using the two  \textit{Chandra} observations using the CIAO tool \texttt{reproject$\_$image}.   Figure~\ref{ima} shows the resultant combined image where  diffuse extended X--ray emission is clearly visible around the compact object. We applied the CIAO \texttt{wavdetect} tool  to the $\sim$90\,ks ACIS-S cleaned image and found RRAT~J1819--1458  at the following position: $\alpha  = 18^{\rm h}19^{\rm m}34.18^{\rm s}$ and  $\delta$ = --14$^{o}$58${'}$03${"}$.7 (error circle of 0$"$.5 radius), in agreement with previous results.

%%%%%%%%%%%%%%%%%%%%%%%    Image             %%%%%%%%%%%%%%%%%%%%%%%%%%%%%%%%%%

\begin{figure}[t]
\begin{center}
\includegraphics[width=\textwidth]{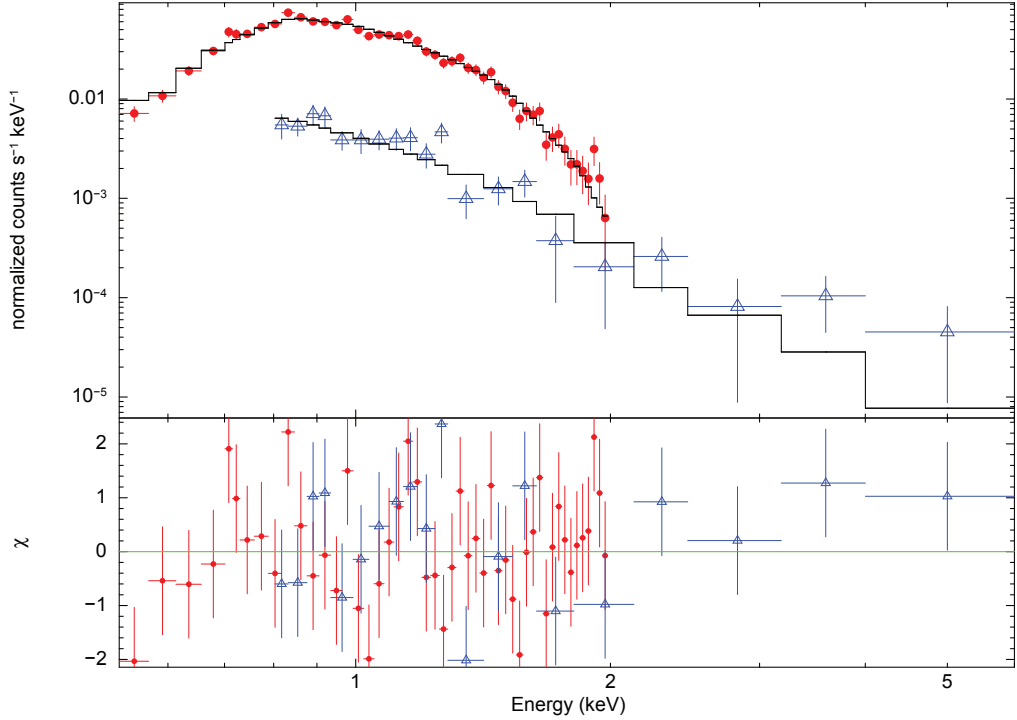}
\end{center}
\caption{Red circles denote the combined ACIS-S spectrum of RRAT J1819--1458$\dag$. Blue open triangles represent the combined spectrum of the extended X--ray emission (Camero--Arranz et al. 2012).}
\label{spec}

\end{figure}
%%%%%%%%%%%%%%%

%\subsection{Timing}

%We used the CIAO tools \texttt{axbary} and \texttt{dmextract}  to create barycentered background-subtracted lightcurves.  The source photons were extracted on each individual observation from a circular region with 2$"$.5 radius (a similar region for the  background but  far from the source was used).  Using the \texttt{Xronos} package, we folded both  X--ray data sets  using the radio ephemeris (\cite[Lyne et al.\ 2009]{lyne09}). The pulse profile shape and  pulsed fraction are all consistent, within the errors, with past measurements (\cite[Reynolds et al.\ 2006]{reynolds06},\cite[McLaughlin et al.\ 2007]{mcLaughlin07},\cite[Rea et al. 2009]{rea09}),  showing no  evidence for long-term variability. 

\subsection{Spectroscopy}

We used the CIAO \texttt{specextract} script to extract source and
background spectra for  RRAT\,J1819--1458. To increase the signal to
noise of the spectrum we proceeded to combine the spectra created for
ObsId 7645 and 12670 using the  CIAO tool \texttt{combine$\_$spectra}
(see Figure 2).  The combined spectrum was modeled  with an absorbed
blackbody plus an absorption line at 1~keV. The Hydrogen absorption
column was fixed  to 0.6$\times$10$^{22}$cm$^{-2}$, allowing us to
better constrain the 1\,keV line feature. The  blackbody temperature
obtained was T$_{BBody}$=0.130$\pm$0.002\,keV,  with
E$_{gauss}$=1.16$\pm$0.03\,keV and $\sigma$=0.17$\pm$0.03\,keV
($\chi_r^2$=1.10; 44 dof). Figure 2 also shows the  $\sim$0.8--7\,keV
combined spectrum for the extended source. An absorbed power law
provides a good fit to the data. The spectral parameters resulted from
the best fitting are $\alpha$=3.7$\pm$0.3 ($\chi_r^2$=1.26; 19 dof).

\subsection{The diffuse X--ray emission structure}
 
To infer the significance and estimate the luminosity of the whole diffuse emission in the combined image, we built the combined Chart/MARX point-spread function (PSF), using both the RRAT J1819--1458 spectrum and its corresponding exposure time. In Figure~3, we compare the surface brightness radial distribution of the combined Chandra observation of RRAT\,J1819--1458~with that of the combined Chart/MARX PSF plus a background level.  This figure shows that the extended emission becomes detectable around 5 pixels ($\sim$2$"$.5) from the peak of the source PSF.  To compute the significance of the diffuse X--ray emission around RRAT\,J1819--1458, from the combined image we extracted all the photons from an annular region of 2$"$.5--20$"$ radii, and we subtracted from it the background extracted from a similar region far from the source.  This resulted in an excess of 790$\pm$18  counts (a detection significance of  $\sim$19$\sigma$).

%%%%%%%%%%%%%%%%%%%%%%%    Image             %%%%%%%%%%%%%%%%%%%%%%%%%%%%%%%%%%

\begin{figure}[t]
\begin{center}
\includegraphics[trim=0 10 0 30,clip, width=0.76\textwidth]{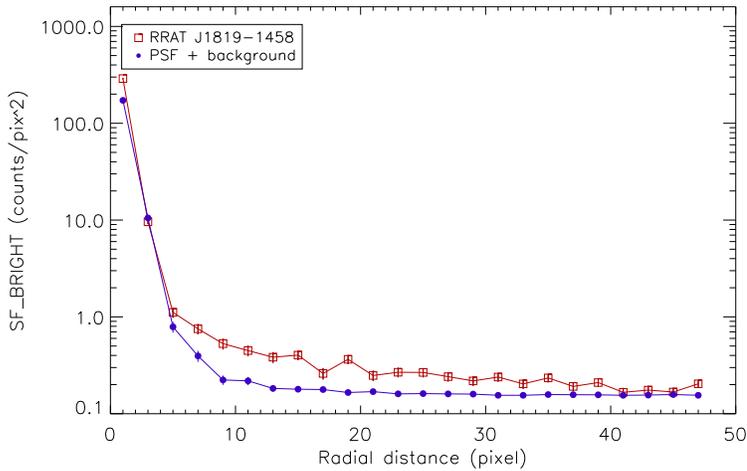}
\caption{Surface brightness of the background-subtracted ACIS-S image
  of RRAT J1819$-$1458 shown in red open squares, and of the Chart/MARX PSF
  plus a constant background shown as blue circles (Camero--Arranz et al. 2012).}
\label{spec}
\end{center}
\end{figure}
%%%%%%%%%%%%%%%

\section{Discussion}

The energies of pulsar wind electrons and positrons range from $\sim$1 GeV to $\sim$1 PeV, placing their synchrotron and inverse
Compton  emission into radio--X--ray and GeV--TeV bands, respectively. This multiwavelength
emission can be seen as a pulsar-wind nebula.  To date, the exact
physical origin and acceleration mechanism of the high-energy
particles in the pulsar  winds are poorly understood, and not all
nebulae can be easily explained as spin-down-powered PWNe. In
\cite[Rea et al.\ (2009)]{rea09} we discussed different scenarios for
the origin of the extended emission detected around RRAT\,J1819--1458.
One option was that the  extended emission we observe is part of the
remnant of the supernova explosion which formed RRAT\,J1819--1458,
unlikely for an object of 117 kyr. A bow-shock nebula due to the
pulsar moving supersonically through the ambient medium was also ruled
out due to  the projected velocity in the case of a bow shock
(v$_p\sim$20\,km\,s$^{-1}$; see Rea et al. 2009 and references
therein)  being rather small.  We propose that \f ~could power a sort
of PWN or the extended X--ray emission around the pulsar might be
explained as a magnetic nebula,  or as a scattering halo as  for
1E\,1547--5408 (Vink \& Bamba 2009, Olausen et al. 2011) and
Swift\,J1834.9-0846 (\cite[Younes et al.\ 2012]{younes12}; Esposito et al. 2012).

\end{document}